\documentclass[twocolumn,showpacs,showkeys,prc,aps]{revtex4}
\usepackage{graphicx}

\newcommand{ \be }{\begin{equation}}
\newcommand{ \ee }{\end{equation}}
\newcommand{ \bea }{\begin{eqnarray}}
\newcommand{ \eea }{\end{eqnarray}}
\newcommand{ \la }{\langle}
\newcommand{ \ra }{\rangle}
\newcommand{ \eps }{{\varepsilon}}

\begin{document}

\title{Methods for the Study of Particle Production Fluctuations}
\author{C. Pruneau} \email{pruneau@physics.wayne.edu}
\author{S. Gavin}    \email{sean@physics.wayne.edu}
\author{S. Voloshin} \email{voloshin@physics.wayne.edu}
\affiliation{Physics and Astronomy Department, Wayne State University, 
666 West Hancock, Detroit, MI 48201}
\date{\today}

\begin{abstract}
We discuss various measures of net charge (conserved quantities)
fluctuations proposed for the identification of critical phenomena in
heavy ion collisions. We show the dynamical component of fluctuations
of the net charge can be expressed simply in terms of integrals of
two- and single-particle densities. We discuss the dependence of the
fluctuation observables on detector acceptance, detection efficiency
and colliding system size and collision centrality.  Finally, we
present a toy model of particle production including charge
conservation and resonance production to gauge the effects of such
resonances and finite acceptance on the net charge fluctuations.

\end{abstract}


\maketitle

\section{Introduction}
\label{sec:intro}

The numbers of particles produced in relativistic nuclear collisions
differ dramatically from collision to collision due to the variation
of impact parameter, energy deposition, baryon stopping and other
dynamical effects \cite{Baym89,Gazdzicki92,Heiselberg01}. Such
fluctuations can also be influenced by novel phenomena such as
disoriented chiral condensate \cite{Rajagopal93,Gavin95} or the
appearance of multiple event classes \cite{Gavin00}. Even globally
conserved quantities such as net charge, baryon number and strangeness
can fluctuate when measured, e.g., in a limited rapidity interval.
The rapid hadronization of a quark-gluon plasma (QGP) can reduce net
charge fluctuations compared to hadronic expectations
\cite{Jeon00,Koch02}, while phase separation can increase net-baryon
fluctuations \cite{Gavin01}. Fluctuations of conserved quantities are
possibly the best probes of such novel dynamics, because conservation
laws limit the degree to which final-state scattering can dissipate
them.

Many statistical measures have been suggested for analyzing particle 
number fluctuations in experiments 
\cite{Mrowczynski02,Jeon00,Heiselberg00,Gavin00}. 
Although these measures superficially appear to be different in nature
and unrelated, closer examination reveals they are in fact connected.
On the other hand, each measure exhibits different dependence on
collision centrality, detector acceptance (rapidity and $p_t$ region
used to calculate the observable), particle detection efficiency, and
susceptibility to experimental biases. The utility of each measure
depends on the particle species measured and the physical phenomena
one wishes to extract. For example, ``robust'' efficiency-independent
measures are best for observing the correlations between neutral and
charged particles produced by disoriented chiral condensate
\cite{Taylor97,Gavin02}.

Experimental efforts to measure event-by-event fluctuations have
followed two approaches. Many advocate a statistical approach in which
fluctuations of particle numbers are characterized by variances,
covariances or other moments
\cite{Mrowczynski02,Jeon00,Heiselberg00,Trainor01,Gavin00,Phenix02,Zaranek01}.
These moments can be compared to expectations based on
thermal-equilibrium or other statistical models; any difference can be
attributed to novel dynamics. Others emphasize the importance of the
momentum-dependent correlation functions, such as the balance
function\cite{Bass00}. The correlation-function approach has yielded
great success in the case of identical pion Hanbury-Brown and Twiss
correlations.

In this paper, we discuss relations between correlation functions and moment 
measures of net-charge fluctuations to study the dependence of these measures 
on collision centrality, experimental efficiency and acceptance. We focus 
initially on the variance $\nu_{dyn}$ suggested in \cite{Voloshin01}, 
which is derived from integrals of the single- and two-particle distribution 
functions. Next, we compare these measures to alternatives suggested in 
\cite{Mrowczynski02,Jeon00}. 
Our correlation-function based analysis complements a study by
Mrowczynski using a statistical point of view
\cite{Mrowczynski02}. Specifically, we begin in
Sect.~\ref{sec:NetCharge} by defining the fluctuation measure
$\nu_{dyn}$ in relation to the microscopic correlation functions. In
the next section, we determine the scaling properties of $\nu_{dyn}$
with system size and, equivalently, collision centrality. We then
introduce alternative fluctuation measures, and discuss their
relationship with $\nu_{dyn}$ in
Sect.~\ref{sec:AlternativeMeasures}. A relation between the net-charge
fluctuations, and the balance function introduced by Bass {\em et
al.}~\cite{Bass00} is presented in
Sect.~\ref{sec:BalanceFunction}. Sect.~\ref{sec:FiniteEfficiency} is
devoted to a discussion on the robustness of fluctuation observables,
i.e., whether and how fluctuation measures introduced in
Sect.\ref{sec:AlternativeMeasures} depend on detection
efficiency. Finally, we consider and compare, in
Sect.\ref{sec:ProductionModels}, the various fluctuation measures in
the context of simple particle production models.

\section{Net-charge fluctuations as a measure 
of 2-particle correlations}
\label{sec:NetCharge}

In this section we show that multiplicity fluctuations are driven by
intrinsic 2-particle correlations. Statistical quantities that we
discuss are constructed from the one-body and two-body densities,
\bea
\rho_1(\eta) &=& \frac{dN}{d\eta},
\nonumber \\
\rho_2(\eta_1,\eta_2) &=& \frac{d^2N}{d\eta_1d\eta_2}.
\label{def} 
\eea
For simplicity, we focus on pseudo-rapidity dependence, although
results can be generalized to address transverse-momentum and
azimuthal-angular dependence.  Our approach and notation in this
section follows Ref.~\cite{Whitmore76,Foa75}.

To extract statistical information from these microscopic densities, we 
use (\ref{def}) to write the multiplicity in the rapidity range 
$\Delta\eta$ as  
\be
\la N \ra = \int_{\Delta\eta}\rho_1(\eta) d\eta.
\label{mean}
\ee
Here $\la N\ra$ represents an average of the observable $N$ over an
event ensemble. Fluctuations of the particle number in this rapidity
range are determined by integrating the two-particle density,
\be
\la N(N-1) \ra = \int_{\Delta\eta} \rho_2(\eta_1,\eta_2) d\eta_1 d\eta_2.
\label{fact2}
\ee
The "-1" appears on the left side because the integral over
$\rho_2(\eta_1,\eta_2)$ counts the average number of particle pairs in
the rapidity interval. Note that $\la N \ra$ and $\la
N(N-1) \ra$ are the first and second order factorial moments of the
multiplicity distribution.

A familiar statistical measure of particle number fluctuations is the
variance,
\be
V = \la (N-\la N \ra)^2 \ra.
\label{variance}
\ee
We can obtain the variance from (\ref{mean}) and (\ref{fact2}), since
$V = \la N(N-1) \ra - \la N \ra(\la N \ra - 1)$. In the absence of
particle-particle correlations, the two-body density factorizes into a
product of two one-body densities. In that case, we find
\be
\la N(N-1) \ra_{\rm uncorr}  
  = \int_{\Delta\eta} \rho_1(\eta_{1})
            \rho_1(\eta_{2}) d\eta_{1}d\eta_{2}  = \la N\ra^2.
\ee
The variance is then $V = \la N \ra$, as expected since the number of
particles produced in a sequence of independent events follows Poisson
statistics \cite{StatisticsBook}. Note that the relative uncertainty
in the mean number $\la N\ra$ is $\sqrt{V}/\la N\ra = 1/\sqrt{\la
N\ra}$ for this case. Observe that the particle number in a grand
canonical ensemble in thermal equilibrium follows Poisson statistics.

Information on net-charge fluctuations is contained in the two-body
density for distinct particles with opposite charges.  We determine
these fluctuations from
\be
\la N_{\alpha}N_{\beta} \ra 
            = \int_{\Delta\eta} \rho_2(\eta_{\alpha},\eta_{\beta}) 
d\eta_{\alpha} d\eta_{\beta},
\label{covariance}
\ee
where $\alpha$ and $\beta$ label the particle species.  In a
statistical framework, this average is related to the two-particle
covariance,
\be
V_{\alpha\beta} = \la N_\alpha N_\beta \ra - \la N_\alpha \ra \la N_\beta\ra.
\label{covariance2}
\ee
The covariance vanishes if there are no correlations between the species 
$\alpha$ and $\beta$, since $\rho_2(\eta_{\alpha},\eta_{\beta})= 
\rho_1(\eta_{\alpha})\rho_1(\eta_{\beta})$.

Following \cite{Whitmore76,Foa75} we define the robust variance,
\be
R_{\alpha\alpha} = {{V-\la N\ra}\over {\la N\ra^2}},
\label{raaCorrelator}
\ee
and the robust covariance, 
\be
R_{\alpha\beta} = {{V_{\alpha\beta}}\over {\la N_{\alpha}\ra \la
N_{\beta}\ra}},
\label{rabCorrelator2}
\ee
for particle species $\alpha$ and $\beta$. These quantities have the
same sensitivity to fluctuations as the variance (\ref{variance}) and
covariance (\ref{covariance2}) but have three significant
advantages. First, these quantities vanish for $V=\la N\ra$ and
$V_{\alpha\beta}=0$, so that they measure the deviation from
Poisson-statistical behavior.  Second -- and of greater practical
importance -- the ratios (\ref{raaCorrelator}) and
(\ref{rabCorrelator2}) are `robust' in that they are independent of
experimental efficiency, To see why Eq.~(\ref{raaCorrelator}) is
robust, let the probability of detecting each
charged particle be $\epsilon$ and the probability of missing it be
$1-\epsilon$. For a binomial distribution the average number of
measured particles is $\langle N\rangle_{\rm exp} =
\epsilon\langle N\rangle$ while the average square is $\langle
N^2\rangle_{\rm exp} = \epsilon^2\langle
N^2\rangle+\epsilon(1-\epsilon)\la N\ra$.
The variance $V_{\rm exp} = \la N^2\ra_{\rm exp}-\la N\ra_{\rm exp}^2 = \epsilon^2 (\la N^2\ra -\la N\ra^2)+\epsilon (1-\epsilon)\la N\ra$, so that
$V_{\rm exp}-\la N\ra_{\rm exp} = \epsilon^2 (V - \la N\ra)$.
We then find 
\begin{equation}
R_{\alpha\alpha}^{\rm exp} = R_{\alpha\alpha},
\end{equation}
independent of $\epsilon$; the proof that (\ref{rabCorrelator2}) is
robust is similar. The ratios (\ref{raaCorrelator}) and
(\ref{rabCorrelator2}) are strictly robust only
if the efficiency $\epsilon$ is independent of multiplicity.
We discuss this point in more detail in section VIII. 

Third, $R_{\alpha\beta}$ are directly related
to the particle correlations. For $\alpha\neq\beta$, we combine (\ref{mean}),
(\ref{covariance}), (\ref{covariance2}) and (\ref{rabCorrelator2}) to obtain
\be
R_{\alpha\beta} = \frac{\int_{\Delta\eta}\rho_2(\eta_{\alpha},\eta_{\beta}) 
d\eta_{\alpha}d\eta_{\beta}}
         {\int_{\Delta\eta}\rho_1(\eta_{\alpha}) d\eta_{\alpha}
          \int_{\Delta\eta}\rho_1(\eta_{\beta}) d\eta_{\beta} }-1;
\label{rabCorrelator}
\ee
one can check that (\ref{rabCorrelator}) also holds for $\alpha = \beta$.
As in an HBT analysis, we define a correlation function $C$ by 
\be
\rho_2(\eta_1,\eta_2) = \rho_1(\eta_1)\rho_1(\eta_2)[1 + C(\eta_1,\eta_2)],
\label{correlatorDef}
\ee
so that (\ref{rabCorrelator}) yields
\be
 R_{\alpha\beta} = \frac{ \int_{\Delta\eta}\rho_1(\eta_{\alpha})\rho_1(\eta_{\beta})
			   C_{\alpha\beta}(\eta_{\alpha},\eta_{\beta}) 
				   d\eta_{\alpha}d\eta_{\beta}}
{\la N_{\alpha}\ra \la N_{\beta}\ra}
\label{rabcab}
\ee
We use this result to illustrate how to extract microscopic
information on the rapidity range of correlations from the
$\Delta\eta$ dependence of $R_{\alpha\beta}$ in sec. VII.

To study net-charge fluctuations, one can measure the robust
covariance for charged hadrons $R_{+-}$. On the other hand, it would
be better to isolate the potentially interesting net-charge
fluctuations from factors that cause the numbers of positive and
negative hadrons to fluctuate together, such as variations in  
energy deposition or collision volume. Toward that end, we
consider dynamic charge observable defined as the linear combination
\be
\nu_{dyn} = R_{++} + R_{--} - 2 R_{+-}.
\label{nuDiff}
\ee
Ratio fluctuations considered by Jeon and Koch \cite{Jeon00} are an
alternative, see sec.~\ref{sec:AlternativeMeasures}. This combination
vanishes when the positive and negative hadrons fluctuate
simultaneously, since all the $R_{\alpha\beta}$ are then the same. We
also see that $\nu_{dyn}$ is both robust (see
sec.\ref{sec:FiniteEfficiency}) and straightforwardly related to the
microscopic correlators (\ref{def}), as are the $R_{\alpha\beta}$. We
find an alternative expression for $\nu_{dyn}$ in terms of
\be
\nu_{+-} = \left\langle \left( \frac{N_+}{\la N_+\ra} - \frac{N_-}{\la N_-\ra}
\right)^2\right\rangle ,
\label{nu}
\ee
where $N_+$ and $N_-$ are respectively the multiplicities of positive
and negative hadrons. In the limit of independent particle
production, $\nu$ becomes
\be
\nu_{stat} = \frac{1}{\la N_+\ra} + \frac{1}{\la N_-\ra}.
\label{nuStat}
\ee
The dynamic charge observable is the difference,
\be
\nu_{dyn} = \nu - \nu_{stat},
\label{nuDynamic}
\ee
as we see by expanding the square in (\ref{nu}).  Observe that
$\nu_{dyn}$ is nonzero when net-charge fluctuations are correlated
(non-Poissonian). Furthermore, eqs.~(\ref{nu}-\ref{nuDynamic}) are
more useful than (\ref{nuDiff}) for extracting correlations from
numerical data since the net-charge fluctuations are typically smaller
than the fluctuations of the total number of hadrons.

We examine the scaling properties of the $\nu_{dyn}$ variance with
collision system size in the next section.

\section{ Scaling of $\nu_{dyn}$  with system size 
and collision centrality in A+A collisions }
\label{sec:Scaling}

We now study the scaling of the observables $C_{\alpha\beta}$,
$R_{\alpha\beta}$, and $\nu_{dyn}$, with collision centrality, target
and projectile mass. For concreteness, we assume that nuclear
collisions are a superposition of independent nucleon-nucleon ($NN$)
sub-collisions and neglect the rescattering of the hadrons. These
assumptions imply that charged-particle pairs can be correlated only if
produced in the same sub-collision. We expect the contribution to the
two-body density from these related pairs to grow linearly with the
number of sub-collisions $M$. Related pairs will be diluted by random
pairs. The $AA$ densities are
\bea
{\rho_{1}}^{AA}(\eta)         &=& M {\rho_{1}}^{NN}(\eta), 
\label{erho1AA}
\\
{\rho_{2}}^{AA}(\eta_1,\eta_2)&=&M {\rho_{2}}^{NN}(\eta_1,\eta_2) \nonumber  \\
                          +M(M &-& 1){\rho_{1}}^{NN}(\eta_1)
				{\rho_{1}}^{NN}(\eta_2).
\label{erho2AA}
\eea
The first term of (\ref{erho2AA}) describes the related pairs while
the second accounts for the $M(M-1)$ random pairs. These expressions
apply generally to particle production from $M$ sources; we focus on
the independent collision model for simplicity. We apply these
considerations to more realistic models at the end of this section.

To compute the correlation function, we substitute the $AA$ densities
Eqs. (\ref{erho1AA}-\ref{erho2AA}) in (\ref{correlatorDef}) to find
\be
C_{\alpha\beta}^{AA}(\eta_1,\eta_2)= 
\frac{C_{\alpha\beta}^{NN}(\eta_1,\eta_2)}{M}.
\label{caaScaling}
\ee
For independent sub-collisions and in the absence of rescattering, we
therefore expect the $AA$ correlation function to have the same
rapidity dependence as in $pp$ collisions, with an overall scale that
is reduced by a factor $M^{-1}$.

Before turning to realistic experiments, we consider for the moment a
collision with a fixed number of sub-collisions. The statistical
observables then satisfy
\be
R_{\alpha\beta}^{AA} = \frac{R_{\alpha\beta}^{NN}}{M} 
\label{rINM}
\ee
and 
\be
\nu_{dyn}(AA) = \frac{\nu_{dyn}(pp)}{M}.
\label{nuINM}
\ee
We see that all quantities scale as $M^{-1}$. 

More realistically, suppose that one specifies a centralility range by
measuring the total charge multiplicity, the zero degree energy, or
some analogous global observable. The number of sub-collisions will
then fluctuate, adding to the variance and covariance of particle
numbers and changing (\ref{rINM}). Specifically, the fluctuations of $M$
contribute a term $\la N_{\alpha}\ra\la N_{\beta}\ra (\la M^2\ra-\la
M\ra^2)$ to the variance and covariance, $V_n$ and $V_{\alpha\beta}$,
so that (\ref{raaCorrelator}) and (\ref{rabCorrelator2}) give
\be
R_{\alpha\beta}^{AA} = \frac{R_{\alpha\beta}^{NN}}{\la M\ra}
+ \frac{\la M^2\ra - \la M\ra^2}{\la M\ra^2}.
\label{RINM2}
\ee
See the appendix for a full derivation. We remark that these $M$
fluctuations are essentially equivalent to the ``volume fluctuations''
discussed in a local equilibrium framework \cite{Jeon00, Koch02}.

On the other hand, random changes in the number of independent
sub-collisions can change the total number of particles but not the
net charge, so that (\ref{nuINM}) is effectively unchanged. We find
\be
\nu_{dyn}(AA) = R_{++}+R_{--}-2R_{+-}=\frac{\nu_{dyn}(pp)}{\la M\ra}.
\label{nuINM2}
\ee
The contributions from sub-collision or volume fluctuations are the
same for all $\alpha$ and $\beta$, so that (\ref{nuDiff}) implies that
this contribution does not affect $v_{dyn}$.  The second term in
(\ref{rINM}) is of order $1/\la M\ra$ and comparable to the first,
since ISR and FNAL experiments suggest that $R_{++}^{NN}\sim
R_{--}^{NN} \sim R_{+-}^{NN}/2$, each of order unity in $\Delta\eta =
1 - 2$ at RHIC \cite{Whitmore76,Foa75}.

We now extend these considerations to the wounded nucleon model, which
sucessfully describes many global features in SPS and AGS
experiments. There, one assumes that only the first sub-collision of
each nucleon drives particle production and neglects all subsequent
interactions \cite{Bialas:1976ed}. Since (\ref{erho1AA}) and
(\ref{erho2AA}) formally describe particle production from $M$
independent sources, we can adapt (\ref{erho1AA}) and (\ref{erho2AA}) to
the wounded nucleon scenario by replacing the number of sub-collisions
$M$ with the number of participant nucleons, $\cal M$. We must also
replace the densities ${\rho_{1}}^{NN}$ and ${\rho_{2}}^{NN}$ in
(\ref{erho1AA}) and (\ref{erho2AA}) with coefficients $\rho_1^0$
and $\rho_2^0$ that describe the production per participant. 
Observe that nucleons are counted as particpants if they interact at 
least once and that there are two participants per $NN$ collision.

Results of the form (\ref{RINM2}) and (\ref{nuINM2}) then follow from
the wounded nucleon model if we replace $M$ with one-half the number
of participants ${\cal M}$. The average number of participants at impact
parameter $b$ for a symmetric $AA$ collision is $\la {\cal M}(b)\ra =2\int
ds T(s)\{1-e^{-\sigma_{NN}T(b-s)}\}$, where $T(b) = \int
\rho(z,b)dz$ is the familiar nuclear thickness function and $\rho$ is
the nuclear density. By comparison, the number of sub-collisions is
$\la M(b)\ra =\sigma_{NN}\int ds T(s)T(b-s)$.  
We remark that both wounded-nucleon and independent-collision 
approximations imply that the total multiplicity of pions 
$N_\pi$ scales as the respective number (participants or subcollisions). 
Therefore, both models imply $\nu_{dyn} \propto N_\pi^{-1}$, albeit 
with different coefficients. 

We point out that particle production at RHIC energy has contributions
from soft interactions, which scale as the number of participants, and
hard processes, which scale as the number of subcollisions
\cite{JeonKapusta,KharzeevNardi}. In this case the scaling of 
$\nu_{dyn}$ with $N_\pi$ can be more complex. Furthermore, final state
scattering effects can certainly modify this scaling.

\section{Alternative Measures of Fluctuations}
\label{sec:AlternativeMeasures}

In this section we consider the connection between 
the variance $\nu_{dyn}$ and other fluctuations measures. 
We discuss some of the merits and problems associated with each observable.

\subsection{$\Phi$-Measure}

The $\Phi$ measure of the net charge fluctuation 
was introduced by Mrowczynski~\cite{Mrowczynski02} and 
is based on statistical considerations. 
It consists of the difference between the mean 
of particle production variances 
calculated event-by-event and the variance calculated over the entire dataset. 
Consider $x$ an observable of interest e.g. the net
charge of produced particles. 
The inclusive mean of $x$ (i.e. average over all particles an events)
is noted $\overline{x}$. Deviation from the inclusive mean are noted
$\Delta x = x - \overline{x}$. 
By construction, one has $\overline{\Delta x}=0$. 
The root mean square (RMS) deviation 
is $\overline{\Delta x^2} = \overline{(x- \overline{x})^2}$.  
To investigate the dynamics, one determines how the event-wise
net value of ``$x$'', defined as $X = \sum_{i} x_i$, changes event by event. 
One defines $\Delta X = X - N \overline{x}$ as the event 
deviation from the inclusive mean 
(with N being the number of particles in the given event). 
By construction, its event average $\la \Delta X \ra$ vanishes, 
whereas $\la \Delta X^2 \ra$ does not. 
The $\Phi$ measure is defined as~\cite{Mrowczynski02}
\be
\Phi = \sqrt{\frac{\la \Delta X^2\ra}{\la N\ra}} - \sqrt{\overline{\Delta x^2}}.
\label{phi}
\ee

For a system with particles of charge $q_+$ and $q_-$, 
the inclusive standard deviation is: 
%
\be
\overline{\Delta x^2} = (q_+ - q_-)^2 \frac{\la N_+\ra \la N_-\ra}{\la N\ra^2}.
\label{meanDeltax2}
\ee
The magnitude of $\la \Delta X^2 \ra$ is determined 
by both statistical and dynamic fluctuations. 
Defining $Q$ as the net charge of an event,
one has $\Delta X=Q-N\la Q\ra/\la N\ra$, from which one 
finds indeed $\la \Delta X\ra = 0$. 
The average $\la \Delta X^2 \ra $ is
however non-zero. 
One finds 
\bea
\la \Delta X^2\ra &=& (q_+ - q_-)^2
\frac{\la N_+\ra^2 \la N_-\ra^2}{\la N\ra^3}  
\nonumber \\
    & & \left( \frac{\la N_+^2\ra-\la N_+\ra^2}{\la N_+\ra^2} \right. 
\nonumber \\
    & & +\frac{\la N_-^2\ra - \la N_-\ra^2}{\la N_-\ra^2} 
\nonumber \\
    & & \left. -2\frac{\la N_+N_-\ra - \la N_+\ra \la N_-\ra}
                      {\la N_+\ra \la N_-\ra} \right),
\label{meanDeltaX2}
\eea
so that 
\bea
 \Phi&=&(q_+ - q_-)\left\{ \frac{\la N_+\ra \la N_-\ra}{\la N\ra^{3/2}} \right.
        \left( \frac{\la N_+^2\ra - \la N_+\ra^2}{\la N_+\ra^2}
 \right. 
\nonumber \\
     & & +\frac{\la N_-^2\ra - \la N_-\ra^2}{\la N_-\ra^2} 
\nonumber \\
     & & \left. -2\frac{\la N_+N_-\ra-\la N_+\ra \la N_-\ra}
                       {\la N_+\ra \la N_-\ra} \right)^{1/2}
\nonumber \\
     & &\left. -\left(\frac{\la N_+\ra\la N_-\ra}{\la N\ra^2} \right)^{1/2}\right\},
\label{phi2}
\eea
Examination of  eqs. (\ref{meanDeltax2}), (\ref{meanDeltaX2}),
and (\ref{phi2}) reveals that they can, in fact, 
be expressed as the $\nu$ and $\nu_{stat}$ variances as follows
(as also reported by Mrowczynski~\cite{Mrowczynski02}):
\bea
\la \Delta X^2\ra &=&  (q_+ - q_-)^2 
\frac{\la N_+\ra^2 \la N_-\ra^2}{\la N\ra^3}
\nu \\
\overline{ \Delta x^2} &=&   (q_+ - q_-)^2 
\frac{\la N_+\ra^2 \la N_-\ra^2}{\la N\ra^3} \nu_{stat} 
\label{meanDeltaX2vsPhi}
\eea
so one can express $\Phi$ as
\be
\Phi = \frac{2 \la N_+ \ra\la N_-\ra}{\la N\ra}\left( \sqrt{\frac{\nu}{\la
N\ra}} 
- \sqrt{\frac{\nu_{stat}}{\la N\ra}}\right).
\label{phiVsNus}
\ee

In general, the dynamic component of the fluctuations is much smaller 
than the statistical component: $\nu_{dyn} << \nu_{stat}$ implying 
$\sqrt{\nu/\la N\ra }-\sqrt{\nu_{stat}/\la N\ra} = \sqrt{\nu_{stat}/\la N\ra}(\sqrt{1+\nu_{dyn}/\nu_{stat}}-1) 
\approx \nu_{dyn}(2\sqrt{\nu_{stat}\la N\ra})^{-1}$. Substituting the value of $\nu_{stat}$ given by Eq.\ref{nuStat}, 
the above expression can thus be approximated by:
\be
\Phi \approx \frac{\la N_+ \ra^{3/2} \la N_-\ra^{3/2}}{\la N\ra^2} \nu_{dyn}.
\label{phiVsNu2}
\ee
One thus finds that indeed the $\Phi$ measure is determined (mostly) 
by the dynamical fluctuations of the system, i.e. by 
the particle correlations implicit in the sum
$R_{++}+R_{--}-2R_{+-}$. 

Eq.~(\ref{phiVsNu2}) further simplifies, as follows, for cases where $\la
N_+\ra=\la N_-\ra$:
\be 
\Phi \approx \frac{\la N\ra}{8} \nu_{dyn}.
\label{phiVsNu3}
\ee
Given, as we discussed in Sect.~\ref{sec:Scaling}, 
that the variance $\nu_{dyn}$ should vary inversely 
to the multiplicity of charge particles in the limit 
of independent particle collisions and absence of 
rescattering of the secondaries, one should expect 
that $\Phi \approx \nu_{dyn,pp}/8$ in that limit, and independent of the
collision centrality if the collision dynamic 
do not vary with collision centrality. 
Note however one must exercise caution while comparing $\Phi$ measured 
by experiments with different 
acceptances (See Sect.~\ref{sec:FiniteAcceptance} for details).
Note finally that unlike $\nu_{dyn}$, the $\Phi$ measure 
is a non robust observable given it explicitly depends 
on the detection efficiency of positive and negative particles 
through the factor $\la N_+\ra$ and $\la N_-\ra$ as we shall 
discuss in more detail in section \ref{sec:FiniteEfficiency}.

\subsection{Particle Ratios}

Another approach advocated in ref.~\cite{Koch02} focuses on the
variance of the ratio of positive and negative particle
multiplicities, $R=\la N_+\ra/\la N_-\ra$.  As shown in
ref.~\cite{Koch02}, the fluctuations of the ratio offer the advantage
that ``volume'' fluctuation effects cancel to first order.  This is
also true for $\nu_{dyn}$ (see sec.~III) and $\Phi$
\cite{Mrowczynski02}.

For small fluctuations, the variance of the ratio can be 
related to the charge variance $\nu$ (\ref{nu}). A small fluctuation of 
$R=\la N_+\ra/\la N_-\ra$ satisfies 
\be
{{\Delta R}\over{R}} = {{\Delta N_+}\over{N_+}} - {{\Delta N_-}\over{N_-}},
\ee
so that
\be
{{\la \Delta R^2\ra}\over{\la R\ra^2}} = 
{{\la\Delta N_+^2\ra}\over{\la N_+\ra^2}} + 
{{\la\Delta N_-^2\ra}\over{\la N_-\ra^2}} -
2{{\la\Delta N_+\Delta N_-\ra}\over{\la N_+\ra\la N_-\ra}}. 
\ee
Expanding the square in (\ref{nu}), we see that 
\be
\la\Delta R^2\ra = \la R\ra^2 \nu
\ee
Observe that neither $\nu$ nor $\la\Delta R^2\ra$ are robust. Also, 
note that this equivalence holds only when $\la \Delta N_{\pm}^2\ra^{1/2} 
<< \la N_{\pm}\ra$; an approximation, which breaks down at small 
multiplicities. Problems with these quantities for small multiplicities
are discussed in \cite{Phenix02}.
The $D$ measure used by Jeon and Koch~\cite{Koch02},
\be
D \equiv\la N\ra\la\Delta R^2\ra = \la N_+ + N_-\ra \la
R\ra^2 \nu,
\label{dRatio}
\ee
is also efficiency dependent.

\subsection{Reduced Variance}

Lastly, we consider the reduced variance $\omega_Q$ used by
authors~\cite{Gavin00,Heiselberg00,Koch02,Bopp02,Phenix02}. If we
write $N = N_+ + N_-$ and $Q=N_+ - N_-$, then the reduced variance is
\be
\omega_{Q} = \frac{\la \Delta Q^2 \ra}{\la N\ra}. 
\label{omega}
\ee
As before, we expand the square to find
\be
\omega_{Q} = 
{{\la\Delta N_+^2\ra + \la\Delta N_-^2\ra -
2\la\Delta N_+\Delta N_-\ra}\over{\la N_+\ra+\la N_-\ra}}. 
\ee
This ratio is unity for Poissonian fluctuations or for a thermal
ensemble in chemical equilibrium; any measured multiplicity dependence
would be interesting.  In terms of robust ratios, we obtain
\be
\omega_{Q} = 1 + 
{{\la N_+\ra^2}\over{\la N\ra}}R_{++}
+{{\la N_-\ra^2}\over{\la N\ra}}R_{--}
-2{{\la N_+\ra^\la N_-\ra}\over{\la N\ra}}R_{+-}.
\ee
Generally, this quantity has a complicated dependence on the correlators $R_{\alpha\beta}$. However,
for $\la N_+\ra\approx \la N_-\ra$, the above expression reduces to
\be
\label{omegaQvsNu}
\omega_{Q} \approx
1 + {{\la N_+ + N_-\ra}\over{4}}\nu_{dyn},
\ee
indicating that this quantity has the same efficiency dependence as the
total number of charged particles.

We note, in closing this section, that the reduced variance, $\omega_Q$, unlike $\nu_{dyn}$, and $\Phi$, has an explicit dependence on collision volume fluctuations, as given by the following expression. 
\be
\omega_Q = \omega_{Q,V} + 
\frac{\left( \la N_+\ra -\la N_-\ra \right)^2}{\la N_+\ra +\la N_-\ra} 
\frac{\la\Delta V^2\ra}{\la V\ra^2}
\ee
where $\omega_{Q,V}$ corresponds to the reduced variance at fixed volume, 
while $\la V\ra$, and $\la \Delta V\ra^2$ are respectively the mean and variance of the collision
volume. The importance of volume fluctuations was pointed out by Jeon and Koch
\cite{Jeon99}. Following their work, it is straightforward to show that $\nu_{dyn}$, and $\Phi$
are independent of volume fluctuations.

\section{Charge conservation effects}

The total charge of the system is fixed due to the charge
conservation. It implies some ``trivial'' correlation in particle
production regardless of other dynamical effects. As such, 
it only affects the two-particle density $\rho_{+-}(\eta_+,\eta_-)$.
We proceed to study the effect of charge conservation on 
the net charge 
fluctuation by calculating the correlation function 
$C_{+-}(\eta_+,\eta_-)$ as a function of single and 
two-particle density expressed in terms of probability distributions 
of positive and negative particle in order to emphasize the role
of charge conservation. One writes, for fixed number $N_{\pm}$ of positive
and negative particles:

\bea
\label{rhoCC1}
\rho_{\pm}(\eta_{\pm})&=& N_{\pm} P_{\pm}(\eta_{\pm}) \\
\label{rhoCC2}
\rho_{+-}(\eta_+,\eta_-)&=&N_{-}P_{+-}(\eta_+,\eta_-)  \\
 & &
 +(N_-N_+-N_-)P_+(\eta_+)P_-(\eta_-)\nonumber
\eea
$P_{\pm}(\eta_{\pm})$ are probabilities to find one + or - particle at
rapidity $\eta_{\pm}$. $P_{+-}$ is the probability to find one
positive particle and one negative particles at rapidities $\eta_+$
and $\eta_-$ respectively.  $N_-$ and $N_+$ are respectively the total
number of negative and positive particles produced (over $4\pi$ solid
angle) by a collision. By virtue of charge conservation, and given the
total charge $Q \geq 0$, one has $Q=N_{+}-N_{-}$, and $N_{+} \geq
N_-$.  The first term of Eq.~(\ref{rhoCC2}) accounts for correlations
between positive and negative particles. As there are $N_-$ +- pairs
created, one has a contribution $N_-P_{+-}$. The second term arises
because there are $N_+N_--N_-$ ways to pair the uncorrelated +-
particles.  In general, at a fixed impact parameter (or number of nn
collisions), the multiplicities $N_-$ and $N_+$ shall fluctuate
event-by-event. One must then average over such fluctuations and
rewrite the above expression as
\bea
\label{rhoCC1a}
\rho_{\pm}(\eta_{\pm})&=& \la N_{\pm}\ra_{4\pi} P_{\pm}(\eta_{\pm}) \\
\label{rhoCC2b}
\rho_{+-}(\eta_+,\eta_-) &=& \la N_{-}\ra_{4\pi} P_{+-}(\eta_+,\eta_-)+  \\
& &(\la N_-N_+ \ra_{4\pi}+\la N_-\ra_{4\pi}) P_+(\eta_+) P_-(\eta_-)\nonumber
\eea
where the notation $\la O \ra_{4\pi}$ represents an average taken over
$4\pi$ acceptance. In the absence of dynamical correlations,
and by virtue of charge conservation, one has
\bea
\la N_-^2\ra_{4\pi} -\la N_-\ra_{4\pi}^2 &=& \la N_-\ra_{4\pi} \\
\la N_-N_+ \ra_{4\pi} &=& \la N_-\ra_{4\pi}^2 +  \la N_-\ra_{4\pi} -  
\la N_-\ra_{4\pi}Q \nonumber
\eea
The correlation function $C_{+-}(\eta_+,\eta_-)$ can then be
calculated and written as
\bea
C_{+-} (\eta_+,\eta_-)
&=&\frac{\rho_{+-} (\eta_+,\eta_-)}{\rho_+(\eta_+)\rho_-(\eta_-)}-1 \nonumber
  \\
&=&\frac{1}{\la N_+\ra_{4\pi}}
   \frac{P_{+-}(\eta_+,\eta_-)}{P_+(\eta_+)P_-(\eta_-)}
\eea

This result is fairly generic and includes the possibility of dynamical 
spatial (or 
rapidity) correlations between the particles of a created pair. Neglecting
such a correlation however, and for the purpose of evaluating the role of 
charge conservation alone, one sets $P_{+-} = P_+P_-$. One then finds that
charge conservation implies:
\be
C_{+-}(\eta_+,\eta_-)=-\frac{1}{\la N_+\ra_{4\pi}}
\approx -\frac{2}{\la N \ra_{4\pi}}
\label{ch_cons_c}
\ee
where $\la N\ra_{4\pi}$ stands for the mean {\em total} number
of charged particles produced in the event. Obviously, at large 
multiplicities one can neglect the difference between $N_+$ and $N/2$.

The correlator $R_{+-}$ is obtained by integration 
(see Eq.~(\ref{rabCorrelator}))
of $C_{+-}(\eta_+,\eta_-)$ over the experimental acceptance. 
Given that $C_{+-}(\eta_+,\eta_-)$ is actually 
independent of $\eta_{\pm}$, $R_{+-}$ is independent of the
experimental acceptance. One thus finds that the charge conservation
contribution to $\nu_{dyn}$ amounts to
\be 
\Delta \nu_{dyn} = -\frac{4}{\la N\ra_{4\pi}}.
\label{eq:nu4pi}
\ee
It is independent of the experimental 
acceptance, and only determined by the total
charge particle multiplicity at a given impact parameter.

We emphasize that $\nu_{dyn}\neq 0$ for a $4\pi$ acceptance because
charge conservation imposes a correlation on the system. The total
$\nu_{+-}$ given by (\ref{nu}) is strictly zero when all particles are
detected. However, (\ref{nuStat}) implies that $\nu_{stat}\neq 0$ in
this case, since the Poisson distributions used to calculate
$\nu_{stat}$ do not incorporate a global charge conservation
constraint. It follows that $\nu_{dyn} = \nu-\nu_{stat} \rightarrow
-\nu_{stat}$ for a $4\pi$ acceptance, as seen in (\ref{eq:nu4pi}).
This estimate of the effect of charge conservation is in agreement
with a correction reported in~\cite{Bleicher00}. Note however that the
correction is additive not multiplicative as stated
in~\cite{Bleicher00}.

\section{Rapidity Dependence of Fluctuations and Detector Acceptance}
\label{sec:FiniteAcceptance}

Measuring the dependence of $R_{\alpha\beta}$ and $\nu_{dyn}$ on the
rapidity window $\Delta\eta$ can yield information on the rapidity
range of correlations as well as their magnitude. Information on the
rapidity dependence of $R_{\alpha\beta}$ is also needed to compare
data from experiments with different geometric acceptance. The
microscopic correlations themselves can and indeed must be determined
from balance function and similar measurements \cite{Bass00}; such
experiments have different practical issues.  We relate $\nu_{dyn}$
and balance function measurements in the next section.

To exhibit the rapidity dependence of $R_{\alpha\beta}$, we assume
that $\rho_1$ are $\eta$ independent and that $C = 
C_{\alpha\beta}(0)\exp\{-(\eta_1 - \eta_2)^2/2\sigma^2\}$. ISR and
FNAL data \cite{Whitmore76,Foa75} show that charged particle
correlation are functions of the relative rapidity $\eta_1-\eta_2$
with only a weak dependence on the average rapidity of the pair. Data
can be roughly characterized as Gaussian near midrapidity.  Using
(\ref{rabcab}) we find
\be
 R_{\alpha\beta} 
 \approx \frac{C_{\alpha\beta}(0)}{x^2}
 \left\{\sqrt{\pi} x\, 
    {\rm erf}(x)- \left(1-e^{-x^2}\right)\right\},
\label{eq:rapDep}
\ee
where $x=\sqrt{2}\Delta\eta/\sigma$. The function $R_{\alpha\beta}$
is shown as a function of $\Delta\eta$ in
Fig. 1. ISR and FNAL data suggest that the rapidity range of
correlations is roughly from 1 to 2 rapidity units.

Both $R$ and the microscopic correlator $C$ depend on the value
$C_{\alpha\beta}(0)$ at $\eta_1 =\eta_2$ and the rapidity range of
correlations, $\sigma$. Equation (\ref{eq:rapDep}) carries the same value
-- and caveats -- as does the Gaussian parameterization of HBT
correlations. The range $\sigma$ depends on the dynamics and may vary with 
centrality as well as target and projectile mass.

One must account for this rapidity dependence when comparing
experiments of different geometrical acceptance. We estimate, for instance, that the
difference between of the fluctuations measured by the STAR
($|\eta|\leq 1.5$,$\Delta\phi=2\pi$) and PHENIX ($|\eta|\leq
0.35$,$\Delta\phi=\pi/2$) experiments to be roughly $\sim$ 10\% for
$\sigma\sim 1-2$. While this is a rather small correction, we
emphasize that the experiments should measure the rapidity dependence.
In general $\sigma$ can differ from $pp$ to $AA$ collisions and,
moreover, is expected to depend on centrality.

\begin{figure}
\centerline{\includegraphics[width=3in]{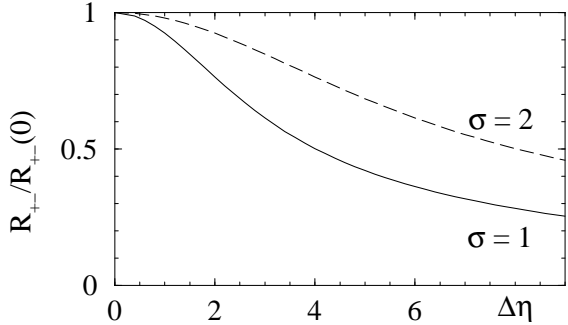}}
\caption[]{Rapidity dependence of the robust covariance $R_{+-}$ assuming 
a Gaussian correlation function of width $\sigma$. }
\label{fig}
\end{figure}

\section{Relation between the balance function and $\nu_{dyn}$}
\label{sec:BalanceFunction}

The balance function was proposed by Bass {\em et al.}~\cite{Bass00}
as a technique to study the dynamics of hadronization in relativistic
heavy ion collisions. The idea is that the rapidity range of
correlations is changed when a collisions forms quark gluon
plasma. Specifically, charged hadrons form late in the reaction,
after hadronization, resulting in shorter-ranged correlations in
rapidity space for charge/anti-charge pairs than expected in the
absence of plasma.

The balance function as defined by Bass {\em et al.}~\cite{Bass00} is written 
(here again focusing, without loss of generality on the rapidity dependence)
\bea
B(\Delta\eta_2|\Delta\eta_1) &=& \frac{1}{2} 
\left\{ D(-,\Delta\eta_2|+,\Delta\eta_1)  \right. \nonumber \\
& & - D(+,\Delta\eta_2|+,\Delta\eta_1)  \nonumber \\
& & + D(+,\Delta\eta_2|-,\Delta\eta_1) \nonumber \\
& & \left. - D(-,\Delta\eta_2|-,\Delta\eta_1) \right\},
\eea
where 
\be
\label{d_fct}
 D(b,\Delta\eta_2|a,\Delta\eta_1)=
\frac{\int_{\eta_1-\Delta\eta_1/2}^{\eta_1+\Delta\eta_1/2}
\int_{\eta_2-\Delta\eta_2/2}^{\eta_2+\Delta\eta_2/2}
d\eta_a d\eta_b \rho_{2}(\eta_b,\eta_a)}
{\int_{\eta_2-\Delta\eta_1/2}^{\eta_2+\Delta\eta_1/2}
d\eta_a \rho_1(\eta_a)}.
\ee
The ratio $D(b,\Delta\eta_2|a,\Delta\eta_1)$ is essentially a 
conditional probability for finding a number of particles of type 
$b$ in the phase space bin $\Delta\eta_2$ of centroid $\eta_2$ given
the presence of particles of type $a$ in the phase space 
bin $\Delta\eta_1$ of centroid $\eta_1$, i.e.
\be
\label{dd}
D(b,\Delta\eta_2|a,\Delta\eta_1)
=\frac{N(b,\Delta\eta_2;a,\Delta\eta_1)}
	{N(a,\Delta\eta_1)}
\ee
The bins need not overlap. 
Experimentally, evaluations of the balance function can be restricted
to a determination of the correlation of particle $a$ and $b$ as a
function of their relative rapidity $\Delta\eta$. In this case,
particle $a$ can be anywhere within the full detector acceptance
$Y$, and particle $b$ is at a rapidity $\Delta\eta$ relative to
$a$. This leads to a one-dimensional balance function,
$B(\Delta\eta|Y)$, defined as:
\bea
B(\Delta\eta|Y) 
&=& \frac{1}{2} \left\{ D(-,\Delta\eta|+,Y)  \right. \nonumber \\
& & - D(+,\Delta\eta|+,Y)  \nonumber \\
& & + D(+,\Delta\eta|-,Y) \nonumber \\
& & \left. - D(-,\Delta\eta|-,Y) \right\}.
\label{pratt}
\eea
To understand this expression better, observe that for a sufficiently 
narrow bin $\Delta\eta$ we can write
\be
D(b,\Delta\eta|a,Y)\approx\frac{\Delta\eta}{\la N_a\ra} 
\int_{-Y/2}^{Y/2} d\eta_a \rho_{2}(\eta_a,\eta), 
\ee
where $\la N_a\ra$ is the number in the full domain 
$-Y/2 \leq \eta \leq Y/2$.
For a boost invariant system the pair correlation function $C$ is a 
function only of the rapidity difference, so that this integral is essentially 
$C$  averaged over the system volume, plus a constant term that cancels in 
(\ref{pratt}). 

The integral of this function over the entire acceptance $Y$ is noted
$B(Y|Y)$. By virtue of Eq.~(\ref{dd}), it amounts to:
\bea
B(Y|Y) &=& \int_{0}^{Y}  d\Delta\eta B(\Delta\eta|Y) 
\nonumber \\
       &=& \frac{1}{2} \left\{\frac{\la N_+ N_-\ra_Y}{\la N_+ \ra_Y }
       \right. \nonumber \\ &&\frac{ \la N_+ N_-\ra_{Y }}{ \la N_-
       \ra_{Y} }
\nonumber \\
  &&\frac{ \la N_+ (N_+-1)\ra_{Y}}{  \la N_+ \ra_{Y} } 
\nonumber \\
  &&\left. 
  \frac{ \la N_- (N_--1)\ra_{Y} } {  \la N_- \ra_{Y} } \right\}
\eea
The four terms of this equation are part of the expression of the
correlators $R_{ab}$ given in
Eqs.~(\ref{raaCorrelator},\ref{rabCorrelator}). The integral $B(Y|Y)$
can thus be re-written:

\bea
B(Y|Y)&=& \frac{1}{2}\left\{ R_{+-}\la N_-\ra + R_{+-}\la N_+\ra \right. \nonumber \\
      & & \left. - R_{++}\la N_+\ra- R_{--}\la N_-\ra \right\} 
\eea
which establishes a relationships between the integral, $B(Y|Y)$, 
of the Balance function,  
and the correlators $R_{++}$, $R_{--}$, and $R_{+-}$.

At RHIC, one observes that 
$\la N_-\ra \approx \la N_+\ra = \la N\ra/2$ near central
rapidities  in Au+Au collisions ~\cite{Phobos01}.  
The above  expression simplifies
\bea
B(Y|Y)& =& \frac{\la N\ra}{4}\left\{ 2R_{+-} - R_{++}-R_{--}\right\} 
\nonumber \\
      & = & -\frac{\la N\ra}{4}\nu_{dyn}.
\eea
The integral, $B(Y|Y)$, of the balance function, $B(\Delta y|Y)$,  
is thus indeed proportional to the variance $\nu_{dyn}$ and 
the total multiplicity $\la N\ra$ 
when $\la N_-\ra \approx \la N_+\ra$. 

\section{Finite Reconstruction Efficiency Effects}
\label{sec:FiniteEfficiency}

We consider the effect of finite reconstruction efficiency on 
measurements of fluctuations studied as a function of collision centrality. 
We assume the centrality is experimentally determined based on the total 
multiplicity of charged particles detected in a reference 
acceptance, $\Omega_M$ whereas the multiplicity fluctuations 
of interest are measured in a fiducial acceptance $\Omega_N$. 
We account for the finite detection efficiency, in a
given acceptance, $\Omega_{\alpha}$, by introducing a detector 
response function $P_{D}(n_{\alpha}|N_{\alpha})$ expressing 
the probability of detecting a multiplicity
$n_{\alpha}$ given a produced multiplicity $N_{\alpha}$. 
In general, $P_{D}(n_{\alpha}|N_{\alpha})$
shall account for finite efficiency effects as well 
as measurements of ghost tracks.
We shall calculate, quite generally, moments, $M_{k,\alpha}$,  
and factorial moments, $F_{k,\alpha}$, of the 
particle multiplicity distribution defined respectively as:
\bea
M_{k,\alpha} &=& \la N_{\alpha}^k \ra = \frac{1}{N_{ev}}\sum N_{\alpha}^k 
\nonumber \\
F_{k,\alpha} &=& \la N_{\alpha}(N_{\alpha}-1) (N_{\alpha}-k)\ra  
\\
             &=& \frac{1}{N_{ev}}\sum N_{\alpha} (N_{\alpha}-1) 
             \cdots (N_{\alpha}-k) ,
\nonumber
\eea
where $N_{ev}$ is the number of events studied. 
The mean  is $\mu_{\alpha}=M_{1,\alpha}$ and the variance, 
$V=\la \delta N_{\alpha}^2 \ra =
M_{2,\alpha}-M_{1,\alpha}^2$. Here we will restrict our calculation 
to these lowest moments, but the calculation can easily be generalized
to higher moments. 

We shall use lower case letter (e.g. $m_{k,\alpha}$) 
to distinguish measured moments from the intrinsic or actual 
moment of the produced particles (i.e. that one wishes to infer) represented
with capital letters (e.g. $M_{k,\alpha}$).

We assume that moments of the multiplicity distributions 
are measured as a function of the collision centrality estimated 
based on the total multiplicity, $m$, measured in the reference acceptance.
The moments can then be expressed (neglecting for simplicity 
the particle type label $\alpha$):
\be
m_{k} = \sum_{n=0}^{\infty} n^k P(n|m),
\ee
where the sum is taken over all relevant multiplicities, 
and $P(n|m)$ is the probability to measure
``n'' given the centrality estimator ``$m$''. 
We emphasize that both ``$n$'' and ``$m$'' are
influenced by the finite efficiency of the detector. 
We in fact seek to extract the intrinsic moments
of the particle production
\be
\label{Mk}
M_{k} = \sum_{N=0}^{\infty} N^k P(N|M),
\ee
where $P(N|M)$ is the probability ``N'' particles are 
produced at a given centrality ``M''.
The measured distribution $P(n|m)$ can be expressed 
as a function of the intrinsic distribution as follows
\be
P(n|m) = \sum_{N,M} P_D(n|N) P(N|M) P_D(M|m),
\ee
with the sum extending over all relevant produced multiplicities $N$ and $M$.
The factor $P_D(M|m)$ corresponds to the probability of having a produced 
multiplicity $M$ given the measured value $m$. 
It is evaluated using Bayes rule:
\be
P_D(M|m) = \frac{P_D(m|M)P(M)}{P(m)},
\ee
where $P(M)$ and $P(m)$ are respectively the probability 
of the produced, M, and measured, m, multiplicities. 
The measured probability distribution is thus
\bea
P(n|m)&=& \frac{1}{P(m)}\sum P(n|N) \nonumber \\
      & & \times P(N|M)P(m|M)P(M) .
\eea

Measured moments can be calculated as function of 
the intrinsic (produced) moment by inserting the above expression 
in (\ref{Mk}). Introducing for convenience the functions $h_s(N)$ 
and $g_s(M)$ defined as follows:
\bea 
h_s(N) &=& \sum_{n} n^s P(n|N) \\
g_s(M) &=& \sum_{N} h_s(N) P(N|M) 
\label{gs}
\eea 
one finds a general expression for the moments as follows:
\be 
\la m_k \ra = \frac{1}{P(m)}\sum_{M} P(m|M)P(M)g_k(M)  .
\ee  
Assuming $P(n|N)$ can be appropriately approximated by 
a binomial distribution, the 
above expressions can be readily simplified. 
The moments $h_s(N)$ yield
\bea
h_1(N) &=& \eps N\nonumber \\
h_2(N) &=& \eps^2 N^2 + \eps (1-\eps) N ,
\eea
where $\eps_n$ is the detection efficiency achieved 
in the measurement of ``n''. 
Substituting these quantities in Eq.~(\ref{gs}) leads to
\bea
  g_1(M) &=& \eps_n \la N \ra
\nonumber \\
  g_2(M) &=& \eps_n^2 \la N^2 \ra + \eps_n(1-\eps_n)\la N \ra .
\eea
 
The first and second moments,  are thus in general 
\bea
\label{expMoments}
\la n\ra  &=& \frac{1}{P(m)} \sum_{M} P_D(m|M) P(M) \eps_n \la N \ra 
\nonumber \\
   \la n^2\ra &=& \frac{1}{P(m)} \sum_{M} P_D(m|M) P(M) 
\nonumber \\
   & &  \times (\eps_n^2 \la N^2\ra+\eps_n(1-\eps_n)\la N\ra)
\eea
with the moments $\la N^s\ra$ evaluated at a fixed value of $M$. 
Clearly, the measured moments are determined by the intrinsic 
moments smeared over the response function of the
multiplicity, $M$. 
Assuming the efficiency of the total multiplicity detection process
is near unity, one can approximate the response 
function $P_D(m|M)$ with a delta function $\delta_{m,M}$, 
and the above expressions simplifies as follows
\bea
\label{simplified}
\la n\ra  &=& \eps_n \la N \ra \nonumber \\
\la n^2\ra& =& \eps_n^2 \la N^2\ra+\eps_n(1-\eps_n)\la N\ra.
\eea

We show in appendix the above results holds for finite efficiency as long 
as ``$n$'' has a linear dependence 
on the total multiplicity ``$m$'' over the range of the response function
$P_D(m|M)$.

We now proceed to use these for the calculation of the various
fluctuation measures introduced in
Sect.~\ref{sec:AlternativeMeasures}. We use sub-indices ``+'', ``-'',
``Q'', and ``CH'' to denote positively and negatively charged
particles, net charge, and total charge particle multiplicity,
respectively.  We use overlined symbols to represent the intrinsic
measures.  We find using Eqs. \ref{simplified}, \ref{phiVsNu2}, \ref{omegaQvsNu}
\bea
   \omega_{\pm} &=& 1-\eps_{\pm} +\eps_{\pm}\overline{\omega}_{\pm} 
\nonumber \\
   \omega_{Q}   &=& 1-\eps_{\pm} +\eps_{\pm}\overline{\omega}_{Q} 
\nonumber \\
    \omega_{CH} &=& 1-\eps_{\pm} +\eps_{\pm}\overline{\omega}_{CH}  
\nonumber \\
   \Phi  &=&  
\frac{\epsilon_+^{3/2}\epsilon_-^{3/2}}{\epsilon^2}\overline{\Phi}.
\eea 

The above fluctuation measures display an explicit dependence 
on the charged particle detection efficiencies 
$\eps_{\pm}$ or the total efficiency $\eps$. 
The $\Phi$ observable, in particular, has a non trivial dependence 
on the detection efficiencies of positively and negatively
charged particles.  
This dependence however simplifies to a single factor, $\eps$, if 
the positive, negative, and global efficiencies 
are equal (i.e. $\eps_+=\eps_-=\eps$).
By contrast, one finds that
the dynamic variance $\nu_{dyn}= \overline{\nu}_{dyn} $ i.e. it 
is independent of the detection efficiencies, and is thus, 
in that sense, a robust observable.  Note that this conclusion
remains strictly correct as long as the Gaussian approximation is 
valid. See the Appendix for a discussion of the Gaussian approximation.

\section{Simple Production Models}
\label{sec:ProductionModels}

\subsection{Poissonian Particle Production}

We first consider a multi-particle production model where no correlation are 
involved. 
Specifically, we assume that on average, particle species, $i$, are 
produced in fixed fractions $f_i$ of the total particle production. 
We consider cases where the fluctuation measures are evaluated 
over kinematic ranges that might be identical (case A) 
or smaller (case B) than the kinematic range used to 
calculate the total (charge) particle production.

The probability to produce species, $i$, with multiplicities, $N_i$, is 
evaluated with a multinomial distribution. In general, one has
\be
P(N_1,N_2, \cdots ,N_k|M) = \frac{1}{M!} \prod_{\alpha=1}^{k} 
            \frac{f_{\alpha}^{N_{\alpha}}}{N_{\alpha}!}  .
\ee
In case A, one shall have 
$M=\sum N_{\alpha} $ and $\sum f_{\alpha} =1$ whereas 
in case B, $M \geq\sum N_{\alpha} $, and $\sum f_{\alpha} <1$.

The multiplicity moments, and variance are calculated at fixed total 
multiplicity, M, assumed to be representative of the collision 
impact parameter:
\bea
   \la N_{\alpha}\ra_m    &=& f_{\alpha} M  
\nonumber \\
   \la N_{\alpha}^2\ra_m  &=& f_{\alpha} M + f_{\alpha}^2 M(M-1) 
\nonumber \\
   \la N_{\alpha}N_{\beta}\ra_m &=& M(M-1)f_{\alpha} f_{\beta} 
\nonumber \\
   \la N_{\alpha}(N_{\alpha}-1)\ra_m &=& f_{\alpha}^2 M(M-1) 
\nonumber \\
   V_{\alpha} &=& M f_{\alpha}(1-f_{\alpha})
\eea
 
Consider now the specific case of net charge fluctuations with the index 
$\alpha$ taking values $+$ and $-$. One has in case ``B'':
\bea
   V_Q         &=& M(f_++f_--(f_+-f_-)^2) 
\nonumber \\
    \omega_Q    &=& 1- \frac{(f_+-f_-)^2}{ f_++f_-} 
\nonumber \\
    \omega_{ch} &=& 1-(f_+-f_-)  
\nonumber \\
     \nu_{+-}    &=& \frac{f_+-f_-}{Mf_+f_-}  
\nonumber \\
    \nu_{dyn}&=&0 
\nonumber\\
     \Phi        &=&0 .
\nonumber\\
\eea
Case A is easily calculated from the above by setting $f_+-f_-=1$.

The coefficients $f_{\pm}$ can be experimentally determined. 
It is thus straightforward to determine the normalized variances 
expected for particle independent production and     
compare with measured values to seek for the presence of 
sub or super Poissonian fluctuations. 
Note additionally that both the $\nu_{dyn}$ and $\Phi$ variables
have null expectation values irrespective of the fraction of 
the fractions $f_{\pm}$.
They thus constitute a more reliable measure of the dynamic fluctuations.

\subsection{Simple Resonance Production Model}
 
Two-particle correlations are determined by a host of phenomena such as 
collective (flow) effects, production of resonances, 
jet production, Fermi/Bose statistics, as well as intrinsic 
phenomena related to the underlying collision dynamics.  
Here we examine the role of resonance decays (e.g. $\rho^o$, $\Delta^o$)
on measurements of the net charge fluctuations. We show that the production
of neutral resonances that decay into pairs of positively 
and negatively charged particles produce an effective dynamical correlation.

We formulate a simple toy model where we include only 
three types of particles: $\pi^+$, $\pi^-$, 
and $\rho^o$. The $\rho^o$ shall be viewed as a generic neutral resonance, 
which decays into $\pi^+$ and $\pi^-$.
Obviously, this is an oversimplification of the problem and a fuller treatment 
shall account for other 
species, all relevant resonances, and the finite acceptance of the detection 
apparatus.  

We consider the $\pi^+$, $\pi^-$, 
and $\rho^o$ to be produced independently (neglecting Bose effects) at 
freeze out in relative fractions $f_1$, $f_2$, and $f_3$ respectively,
and model the multiplicity production according to 
a multinomial distribution (as in the previous section). 
The probability of producing 
$n_1$ $\pi^+$, $n_2$ $\pi^-$, and $n_3$ $\rho^o$ is expressed
\be
P(n_1,n_2,n_3;N)=\frac{N!}{n_1!n_2!n_3!}f_1^{n_1}f_2^{n_2}f_3^{n_3} .
\ee
Given our assumption that all  $\rho^o$ decay into 
a pair $\pi^+$ and $\pi^-$, 
the probability of measuring $n_+$ positive $n_-$ negative particles 
respectively can be written
\bea
    P(n_+,n_-;N)&=& \sum_{n_1,n_2,n_3}P(n_1,n_2,n_3;N) 
\nonumber \\
    & & \times \delta_{n_+,n_1+n_3}\delta_{n_-,n_2+n_3}.
\eea
One then writes the moment generating function of 
the probability  $ P(n_+,n_-;N)$ as
\be
G(t_+,t_-;N)=(p_1 e^{t_+}+p_2 e^{t_-}+p_3e^{t_++t_-} )^N ,
\ee 
which one uses to computes the moments of the pion multiplicity distributions.
One finds:
\bea
   \la N_+\ra &=& N(f_1+f_3) 
\nonumber \\
    \la N_-\ra &=& N(f_2+f_3) 
\nonumber \\
    \la N_+(N_+-1)\ra &=& N(N-1)(f_1+f_3)^2 
\nonumber \\
    \la N_-(N_--1)\ra &=& N(N-1)(f_2+f_3)^2 
\nonumber \\
   \la N_+N_ \ra &=& N(N-1)(f_1+f_3)(f_2+f_3)
\nonumber \\
  & & +Nf_3 .
\eea
The variance $\nu_{dyn}$, in the presence of resonances, is thus 
simply
\be
\nu_{dyn}=\frac{-2p_3}{N(p_1+p_3)(p_2+p_3)}.
\ee

One finds that  the variance $\nu_{dyn}$ increases with 
the fraction of resonances, $p_3$ produced in the final state. 
One also finds it to scale inversely to the number
of particles produced in the initial state. 
Note that in the limit $p_3=0$, $\nu_{dyn}$
vanishes by our assumption of independent production. 
The simple treatment done here does not account for finite acceptance 
effects on the decay of resonances. 
Obviously, if too small a rapidity region is 
integrated, one of the decay partners may on average be missed, 
and $|\nu_{dyn}|$  shall be increased accordingly.

In AA collisions, one does not expect resonance production 
to be the sole cause of correlation, i.e. $\nu_{dyn}<0$, 
but it is yet to be determined what fraction
of the observed fluctuations may be attributed to resonance 
production or to truly dynamic correlations. 
In that respect, it shall be interesting to consider fluctuations 
of specific particle species such  $p/\overline{p}$
  in contrast to $\pi^{\pm}$  or $K^{\pm}$ given no known resonance 
decay into $p+\overline{p}$ whereas many 
resonances exist that decay into  $\pi^++\pi^-$ or $K^++K^-$.

\section{ Summary and Conclusions}
\label{sec:conclusions}

We introduced the net charge fluctuation measure, $\nu_{dyn}$, on the basis of 
two-particle correlation functions. We showed that for heavy ion collisions
involving independent nucleon collision and negligible rescattering of
secondaries, $\nu_{dyn}$ scales as the multiplicative inverse of the produced 
charged particle multiplicity. We also showed that $\nu_{dyn}$ 
is simply related to other observables used or proposed for 
fluctuation measurement by various 
authors. We found however that the different fluctuation measures 
have different dependence on the experimental acceptance, detector
efficiency, and collision centrality.  We showed that $\nu_{dyn}$ has 
a weak dependence on the 
rapidity range used experimentally to measure the fluctuations provided 
the rapidity range is of the order or smaller than the two-particle 
correlator width, whereas observables such as $\Phi$ have basically
a linear dependence on the size of the acceptance. We found also that
$\nu_{dyn}$ is, by construction, independent, to first order, 
of the detection efficiency whereas measures such as $\Phi$, $\omega_Q$
have a explicit dependence on the detection efficiency. We also found
that charge conservation has a finite, and actually 
sizable  effect on the charge fluctuation measure $\nu_{dyn}$
determined by the total charge particle multiplicity (over $4\pi$
and independent of the detector acceptance used to measure the
net charge fluctuations.
We further showed, as also pointed out by Mrowczynski~\cite{Mrowczynski02}
, that the $\Phi$ measure shall be 
independent of the collision centrality provided the collision dynamic 
is also independent of the collision centrality. Note however that because
the detection efficiency may be a subtle function of the detector occupancy, and
hence the collision centrality, caution has to be exercised when interpreting
uncorrected measurement of $\Phi$ vs. collision centrality.
Finally, we presented, as an example, a simple particle production model that can be
used to account for the production of resonances as well as charge
conservation. 

\acknowledgments{
We thank R. Bellwied
for discussions. This work 
was supported in part by U.S. DOE Grant No. DE-FG02-92ER40713.
}

\appendix
\section{Finite Efficiency Effects on the Measurement of $\nu_{dyn}$}

A range of collision impact parameters is selected in experiments
using a measured multiplicity $m$ (or a similar observable). This
introduces additional fluctuations because a single $m$ corresponds to
a range of impact parameters. In this appendix we estimate the effect of 
centrality selection. We use these results in secs.~\ref{sec:Scaling} and 
\ref{sec:FiniteEfficiency}. 

We assume, in the Gaussian approximation, that the moments scale with the true
multiplicity $M$ as
\bea
\label{linearMoments}
\la N_a\ra &=&\mu_a M, \nonumber \\
\la N_a^2\ra &=& \mu_a^2 M^2 + \sigma_a^2 M, \nonumber \\
\la N_aN_b\ra &=& \mu_a\mu_b M^2 +\xi_{ab}M 
\eea
where $\mu_a$ and $\mu_b$ are average branching fractions for the production of
 species ``a'' and ``b'' respectively, while $\sigma_a^2$ and $\xi_{ab}$ are their variance and covariance.
These relations are strictly true in the independent collision model
or the wounded nucleon model where both $M$ and $N_a$ are respectively
proportional to the number of sub-collisions or the number or strings.
The first moment (\ref{expMoments}) is then
\bea
\la n_a\ra  &=& \frac{1}{P(m)} \sum_{M} P_D(m|M) P(M) \eps_a \mu_a M \nonumber \\
 &=&  \eps_a \mu_a  \frac{1}{P(m)} \sum_{M}M P_D(m|M) P(M) \nonumber \\
 &=&  \eps_a \mu_a \la M\ra_m
\eea
where we have introduced the expectation value of $M$ at fixed $m$ defined as
\be
\la M\ra_m = \frac{1}{P(m)}\sum_{M}M P_D(m|M) P(M)
\ee
The factor $\epsilon_a$ is the probability that a particle of type ``a'' is detected. 
One gets similarly for the 2nd moment and cross term:
\bea
\la n_a^2\ra  &=& \frac{1}{P(m)} \sum_{M} P_D(m|M) P(M) \left[\eps_a^2(\mu_a^2M^2+\sigma_a^2M) \right.  \nonumber \\
 & & \left. +\eps_a(1-\eps_a)\mu_a M \right], \nonumber \\
\la n_a^2\ra  &=& \left[ \eps_a^2 (\sigma_a^2-1)+\eps_a\right]\la M\ra_m + \eps_a^2\mu_a^2\la M^2\ra_m, \nonumber \\
\eea
and
\bea
\la n_a n_b\ra &=& \eps_a\eps_b \mu_a\mu_b\left(\la M^2\ra_m-\la M\ra_m^2\right)\nonumber \\
    & & + \eps_a\eps_b\xi_{ab}\la M\ra_m.
\eea
The correlators $R_{aa}$ and $R_{ab}$ are therefore
\be
R_{aa} =\frac{\sigma_a^2-\mu_a}{\mu_a^2}\frac{1}{\la M\ra_m} +
\frac{\la M^2\ra_m-\la M\ra_m^2}{\la M\ra_m^2} 
\ee
and
\be
R_{ab} =\frac{\xi_{ab}}{\mu_a^2}\frac{1}{\la M\ra_m} +
\frac{\la M^2\ra_m-\la M\ra_m^2}{\la M\ra_m^2} 
\ee
The variance $\nu_{dyn} = R_{aa}+R_{bb}-2R_{ab}$ measured at a given $m$ is then 
\be
\nu_{dyn}(m) 
 = \frac{\nu_0}{\la M\ra_m}
\ee
where
\be
\nu_0 = \frac{\sigma_a^2-\mu_a}{\mu_a^2} 
 +  \frac{\sigma_b^2-\mu_b}{\mu_b^2} 
 - 2 \frac{\xi_{ab}}{\mu_a\mu_b}.
\ee
This expression amounts to the value of $\nu_{dyn}$ evaluated at
$M=\la M\ra_m$. One finds that the correlators $R_{ab}$ exhibit a
contribution from the variance $\la M^2\ra_m-\la M\ra_m^2$ whose
magnitude depends on the detector response function width.  The
variance $\nu_{dyn}$ however does not have such a contribution and as
such is also independent of the detection efficiency for measuring
$M$.

Note that the above result implies that $\nu_{dyn}$ is robust, i.e. independent 
of detection efficiencies, in the Gaussian approximation (\ref{linearMoments}). 
An explicit dependence on efficiencies would arise
if the Gaussian approximation is not valid, e.g. if the detector 
response functions differ markedly from Binomial or Gaussian functions, or if the 
efficiencies exhibit very large variations with detector occupency.

\bibliography{method}

\begin{thebibliography}{10}

\bibitem{Baym89}
  Gordon Baym, Gerald Friedman and Ina Sarcevic, Phys.\ Lett.\ {\bf B219},
  (1989) 205.

\bibitem{Gazdzicki92}
M. Gazdzicki and S.\ Mrowczynski, Z.\ Phys.\ {\bf C54} (1992) 127.

\bibitem{Heiselberg01}
H. Heiselberg, Phys.\ Rept.\ {\bf 351} (2001) 161.

\bibitem{Rajagopal93}
K. Rajagopal and F. Wilczek, Nucl. Phys. B399, (1993)395.

\bibitem{Gavin95}
S. Gavin, Nucl. Phys. A590 (1995) 163c.

\bibitem{Gavin00}
{\em Covariance of anti-proton yield and source size in nuclear collisions } \\
  S. Gavin and C. Pruneau, Phys. Rev. C61 (2000) 044901.

\bibitem{Jeon00}
  S. Jeon, V. Koch, Phys. Rev. Lett. 85, 2076 (2000).

\bibitem{Koch02}
V. Koch, M. Bleicher, S. Jeon, Nucl.Phys. A698 (2002) 261-268.

\bibitem{Gavin01}
D. Bower and S. Gavin, Phys. Rev. C64 (2001) 051902(R).

\bibitem{Mrowczynski02}
S. Mrowczynski, nucl-th/0112007.

\bibitem{Heiselberg00}
H. Heiselberg and A. D. Jackson, Phys. Rev. C63 (2001) 064904.

\bibitem{Taylor97}
T.~C.~Brooks {\em et al.} Phys. Rev. D55, (1997) 5667; Phys. Rev. D61,(2000)
  032003.

\bibitem{Gavin02}
  S. Gavin and J.~I.~Kapusta, Phys. Rev. C (2002) in press.

\bibitem{Trainor01}
T.~A.~Trainor (STAR Collaboration), Proc. 17th Winter Workshop Nuclear Dynamics (2001).

\bibitem{Phenix02}
  K.~Adcox, {\em et al.} (PHENIX Collaboration), NUCL-EX/0203014.

\bibitem{Zaranek01}
J. Zaranek,  hep-ph/0111228.

\bibitem{Bass00}
S. Bass, P. Danielewicz, and S. Pratt, Phys. Rev. Lett. 85,
  2689 (2000).

\bibitem{Voloshin01}
S. Voloshin {\em et al.}, Proc of INPC2001, Berkeley, CA 2001.

\bibitem{Whitmore76}
J. Whitmore, Physics Reports 27, 187 (1976).

\bibitem{Foa75}
L. Foa, Physics Reports 22, 1 (1975).

\bibitem{StatisticsBook}
Glen Cowan, Statistical Data Analysis, Oxford Science Publications (1998).

\bibitem{Bialas:1976ed}
A.~Bialas, M.~Bleszynski and W.~Czyz, Nucl.\ Phys.\ B {\bf
  111} (1976) 461.

\bibitem{JeonKapusta}
S. Jeon and J.~I. Kapusta, Phys. Rev. C63 (2001) 011901.

\bibitem{KharzeevNardi}
D. Kharzeev and M. Nardi, Phys. Lett. B507 (2001) 121.

\bibitem{Bopp02}
F.~W.~Bopp and J.~Ranft, hep-ph/0105192.

\bibitem{Jeon99}
S. Jeon, V. Koch, Phys. Rev. Lett. 83, 5435 (1999).

\bibitem{Bleicher00}
M. Bleicher, S. Jeon, V. Koch, Phys.Rev. C62 (2000) 061902.

\bibitem{Phobos01}
B.~Back {\em et al.} (PHOBOS Collaboration), Phys. Rev. Lett. 87 (2001) 102301.

\end{thebibliography}
\bibliographystyle{unsrt}    

\end{document}